\documentclass[fleqn,pra]{revtex4}

\usepackage{amsmath,graphicx}
\usepackage{dcolumn}
\usepackage{verbatim}

\begin{document}

\title{Rovibrational spin-averaged transitions in the hydrogen molecular ions}

\author{Vladimir I. Korobov}
\affiliation{Bogoliubov Laboratory of Theoretical Physics, Joint Institute for Nuclear Research, Dubna 141980, Russia}
\author{J.-Ph.~Karr}
\affiliation{Laboratoire Kastler Brossel, Sorbonne Universit\'e, CNRS, ENS-Universit\'e PSL, Coll\`ege de France, 4 place Jussieu, F-75005 Paris, France}
\affiliation{Universit\'e d'Evry-Val d'Essonne, Universit\'e Paris-Saclay, Boulevard Fran\c cois Mitterrand, F-91000 Evry, France}

\begin{abstract}
We reconsider the calculation of rovibrational transition frequencies in hydrogen molecular ions. Some previously neglected contributions, such as the deuteron polarizability, are included into consideration in comparison with our previous work. In particular, one-loop and two-loop QED corrections at $m\alpha^7$ and $m\alpha^8$ orders are recalculated in the framework of the adiabatic approximation, with systematic inclusion of corrections associated with vibrational motion. Improved theoretical transitions frequencies are obtained and found to be in very good agreement with recent high-precision spectroscopy experiments in HD$^+$. New values for the $m_p/m_e$ and $m_d/m_p$ mass ratios are determined.
\end{abstract}

\maketitle

\section{Introduction}

Considerable progress has been recently achieved in the spectroscopy of hydrogen molecular ions. Several rovibrational transitions in HD$^+$ have been measured with relative uncertainties in the 10$^{-11}$--10$^{-12}$ range through Doppler-free spectroscopy of ultracold trapped ions in the Lamb-Dicke regime~\cite{Schiller20,Patra20,Kortunov21}, approaching or even exceeding the precision of theoretical predictions~\cite{Korobov17}. This has allowed to get improved determinations of the proton-to-electron mass ratio, and to perform tests of QED constraining hypothetical ``fifth forces'' between hadrons~\cite{Schiller20,Germann21}. These results, as well as other ongoing projects~\cite{Zhong15,Schmidt20}, and perspectives of reaching higher precision~\cite{Schiller14,Karr14} e.g. using quantum-logic spectroscopy schemes~\cite{Schmidt05,Wolf16,Chou20} provide strong motivation to improve the theory further.

In our paper~\cite{Korobov17}, we calculated the frequencies of fundamental vibrational transitions in the nonrelativistic QED approach, including corrections up to the $m\alpha^8$ order. Since then, several new advances in the theory of hydrogenlike atoms have been achieved~\cite{YPP19,Szafron19,Karshenboim19}, which allows to get improved results for the corresponding correction terms in the hydrogen molecular ions. In addition, we found that several contributions that had been neglected in our previous consideration were of comparable magnitude to the estimated error bar, and thus should be included. Finally, changes in the recommended values of fundamental constants, the nucleus-to-electron mass ratios but also the Rydberg constant, proton and deuteron radii, between the previous (2014) CODATA adjustment~\cite{CODATA14} and the most recent one (2018)~\cite{CODATA18} (see Table~\ref{tab:CODATA-fc}) also affect our theoretical predictions and their uncertainties.

The aim of this work is to reanalyse the theory, with particular emphasis on the evaluation of QED corrections at orders $m\alpha^7$ and $m\alpha^8$ in the framework of the adiabatic approximation, where we systematically include ``vibrational'' corrections, i.e. the second-order perturbation terms due to perturbation of the vibrational wavefunction. Improved theoretical rovibrational transition frequencies are given for experimentally relevant transitions, and the impact of these new results on the determination of the proton-to-electron mass ratio is illustrated.

We use atomic units throughout this paper ($\hbar\!=\!m_e\!=\!e\!=\!1$). Other constants used in calculations are taken from the CODATA18 adjustment~\cite{CODATA18} (see in Table~\ref{tab:CODATA-fc}), including the fine structure constant, $\alpha=7.297\,352\,5693(11)\!\times\!10^{-3}$.

\section{Nuclear size and polarizability corrections} \label{sec2}

In our previous calculations~\cite{Korobov17,Korobov06}, we only included the leading-order nuclear finite-size correction, see Eq.~(6) in~\cite{Korobov06}. Some higher-order nuclear corrections are not negligible at the current level of theoretical accuracy, in particular the deuteron polarizability~\cite{Friar97}. Here, we follow the notations used in~\cite{CODATA14}. According to Eq.~(34) of~\cite{CODATA14}), we write the $m\alpha^5$ deuteron polarizability contribution as
\begin{equation}
E_{\rm pol}^{(5)}(\hbox{D}) = [-21.37(8)]\, \left\langle\pi\delta(\mathbf{r}_d)\right\rangle\, (\hbox{kHz}).
\end{equation}
For example, this results in a 0.33~kHz shift for the frequency of the fundamental vibrational transition $(L=0,v=0)\to(0,1)$, which is comparable to the overall theoretical uncertainty of 0.5~kHz for this transition (see Table~\ref{ftE}).

Nuclear finite-size corrections at the same order~\cite{Friar97b} are written as (see Eq.~(59) in~\cite{CODATA14})
\begin{equation}
E_{\rm fns}^{(5)}(\hbox{D}) = -(2R_{\infty}c)\frac{2\pi}{3}C_{\eta}\left(\frac{R_d}{a_0}\right)^3\,\left\langle\pi\delta(\mathbf{r}_d)\right\rangle
 = [-0.57(3)]\, \left\langle\pi\delta(\mathbf{r}_d)\right\rangle\, (\hbox{kHz}),
\end{equation}
where $a_0$ is the Bohr radius, $R_d=2.12799(74)$ the rms charge radius of the deuteron, and $C_{\eta}=2.0(1)$.

The contribution at the next order ($m\alpha^6$), unlike the previous ones, is not a ``state-independent'' term proportional to the squared value of the wavefunction at the electron-nucleus, and would require an independent calculation. However, this term can be estimated from its value for the hydrogen atom ground state by using the LCAO approximation for the electronic wavefunction:
\[
\psi_{_{\rm LCAO}}^{}(\mathbf{r}) = \frac{1}{\sqrt{2}}\bigl(\psi_{1s}(\mathbf{r}_p)\!+\!\psi_{1s}(\mathbf{r}_d)\bigr),
\]
where $\psi_{1s}$ is the hydrogen ground state wavefunction. Under this approximation, one gets from Eq.~(59) of~\cite{CODATA14}:
\begin{equation}
E_{\rm fns}^{(6)}(\hbox{D})
   = -(2R_{\infty}c)\frac{2\pi}{3}\left(\frac{R_d}{a_0}\right)^2
      (Z\alpha)^2\left(C_{\theta}-\ln\frac{ZR_d}{a_0}\right)\,\left\langle\pi\delta(\mathbf{r}_d)\right\rangle
   = [3.96(2)]\, \left\langle\pi\delta(\mathbf{r}_d)\right\rangle\, (\hbox{kHz}).
\end{equation}
with $C_{\theta}=0.38(4)$. Only the uncertainty from $C_{\theta}$ is indicated in this equation. Due to the employed LCAO approximation, one may estimate the uncertainty of $E_{\rm fns}^{(6)}(\hbox{D})$ as equal to the nonlogarithmic term, which is still much smaller than the overall theoretical uncertainty.

In the proton case, all the corrections considered above are negligibly small at the current level of accuracy.

\begin{table}[t]
\begin{center}
\begin{tabular}{l@{\hspace{2mm}}l@{\hspace{2mm}}l@{\hspace{2mm}}l}
\hline\hline
quantity & symbol & value & uncertainty \\
\hline
\vrule width 0pt height 11pt
proton charge radius & $r_p$ & $0.8751(61)\!\times\!10^{-15}\mbox{ m}$ & $7.0\!\times\!10^{-3}$
\\
 & & $0.8414(19)\!\times\!10^{-15}\mbox{ m}$ & $2.2\!\times\!10^{-3}$
\\
Rydberg constant & $R_\infty=\alpha^2m_ec/2h$ & $10\,973\,731.568\,508(65) \mbox{ m}^{-1}$ & $5.9\!\times\!10^{-12}$
\\
 &  & $10\,973\,731.568\,160(21) \mbox{ m}^{-1}$ & $1.9\!\times\!10^{-12}$
\\
proton-to-electron  & $\mu_p=m_p/m_e$ & $1836.152\,673\,89(17)$ & $9.5\!\times\!10^{-11}$
\\
mass ratio          & & $1836.152\,673\,43(11)$ & $6.0\!\times\!10^{-11}$
\\
deuteron-to-electron  & $\mu_d=m_d/m_e$ & $3670.482\,967\,85(13)$ & $3.5\!\times\!10^{-11}$
\\
mass ratio          & & $3670.482\,967\,88(13)$ & $3.5\!\times\!10^{-11}$
\\
\hline\hline
\end{tabular}
\end{center}
\vspace*{-5mm}
\caption{Reevaluation of the fundamental constants of atomic physics by CODATA in 2018. CODATA14 (resp. CODATA18) values are given in the upper (resp. lower) line.}
\label{tab:CODATA-fc}
\end{table}

\section{$m\alpha^7$ and $m\alpha^8$-order corrections in the adiabatic approximation}\label{sec3}

Relativistic and QED corrections at the orders $m\alpha^4$ to $m\alpha^6$ have been evaluated in a full three-body approach using precise variational wavefunctions~\cite{Korobov06,Korobov12}, except for the $m\alpha^6$ relativistic correction~\cite{Korobov17,KorobovJPB07}. For calculation of $m\alpha^7$ and higher order one- and two-loop corrections we use the Born-Oppenheimer approach, where the states of the molecule are taken in the form
\begin{equation}\label{BO}
\Psi^{\rm BO} = \phi_{\rm el}(\mathbf{r};R)\chi_{\rm BO}^{}(R).
\end{equation}
The electronic wavefunction obeys the clamped nuclei Schr\"odinger equation for a bound electron
\begin{equation}\label{BO_el}
\left[H_{\rm el}-\mathcal{E}_{\rm el}(R)\right]\phi_{\rm el}=0,
\end{equation}
where
\[
H_{\rm el} = \frac{p^2}{2m} + V + \frac{Z_1Z_2}{R},
\qquad
V = -\frac{Z_1}{r_1} - \frac{Z_2}{r_2}.
\]
Here $H_{\rm el}$ is the electronic Hamiltonian, $Z_1$ and $Z_2$ are the charges of the nuclei, and $r_1$, $r_2$ are the distances from the electron to nuclei 1 and 2, respectively. The wavefunction $\chi_{\rm BO}^{}(R)$ describes the relative nuclear motion, and is a solution of
\begin{equation}\label{radial}
\left(H_{vb}\!-\!E_0\right)\chi_{\rm BO}^{}
 = \left[-\frac{\nabla_R^2}{2\mu_N}\!+\!\mathcal{E}_{\rm el}(R)\!-\!E_0\right]\chi_{\rm BO}^{}
 = 0,
\end{equation}
where $\mu_N=M_1M_2/(M_1\!+\!M_2)$ is the reduced mass of the nuclei.

Instead of the Born-Oppenheimer solution $\chi_{\rm BO}^{}(R)$ we use the adiabatic solution $\chi_{\rm ad}(R)$, which includes as well the adiabatic corrections
\begin{equation}
\mathcal{E}_{\rm ad}(R) = \mathcal{E}_{\rm el}
   +\int d\mathbf{r}
   \left\langle
      \phi_{\rm el}\left|
         \frac{\mathbf{p}^2}{8\mu_N}+\frac{\mathbf{P}^2}{2\mu_N}-\frac{\kappa}{2\mu_N}\mathbf{p}\mathbf{P}
      \right|\phi_{\rm el}
   \right\rangle
\end{equation}
where $\mathbf{p}$ is the electron impulse in the center-of-mass frame, $\mathbf{P}$ the relative impulse of the two nuclei, and $\kappa = (M_1\!-\!M_2)/(M_1\!+\!M_2)$ is the asymmetry parameter. See Ref.~\cite{Wolniewicz80}, or a review by Carrington {\em et al.} \cite{Carrington89} for more details.

The one-loop self-energy correction of order $m\alpha^7$ to the energy of a bound electron in the two-center problem (non-recoil limit) has been determined in~\cite{JCP_PRL05,JCP_PRA05,Korobov13,KorobovPRL14,KorobovPRA14}. The electronic part of the correction can be calculated using the effective Hamiltonian of Eq.~(6) from Ref.~\cite{KorobovPRA14}:
\begin{equation}\label{EffH}
\begin{array}{@{}l}
\displaystyle
\Delta E_{\rm el-SE}^{(7)} =
   \left\langle
      \chi_{\rm ad}\bigl|\mathcal{E}_{\rm 1loop-SE}^{(7)}(R)\bigr|\chi_{\rm ad}
   \right\rangle,
\end{array}
\end{equation}
numerical data for the $\mathcal{E}_{\rm 1loop-SE}^{(7)}(R)$ effective potential may be found in the Supplemental Material to \cite{Korobov13}.

The one-loop vacuum polarization (Uehling potential) contribution has been considered in~\cite{KarrVP17}. The adiabatic approximation was also compared with full three-body calculations, confirming that it is accurate to $\mathcal{O}(m/M)$ (where $m/M$ is the electron-to-nucleus mass ratio). The electronic part of the correction can be written as
\begin{equation}\label{EffH}
\begin{array}{@{}l}
\displaystyle
\Delta E_{\rm el-VP}^{(7+)} =
   \left\langle
      \chi_{\rm ad}\bigl|\mathcal{E}_{\rm 1loop-VP}^{(7+)}(R)\bigr|\chi_{\rm ad}
   \right\rangle,
\end{array}
\end{equation}
where $\mathcal{E}_{\rm 1loop-VP}^{(7+)}(R)$ is given by Eq.~(16) of~\cite{KarrVP17}.

As shown in~\cite{Korobov17,KarrVP17}, beyond the above electronic contributions one also needs to include vibrational corrections. The latter are second-order perturbation terms stemming from perturbation of the vibrational wavefunction by the leading relativistic and radiative corrections to the adiabatic potential $\mathcal{E}_{\rm ad}(R)$, namely
\begin{equation} \label{leading}
\begin{array}{@{}l}\displaystyle
\mathcal{E}_{\rm BP}^{(4)}(R) =
   \alpha^2\Bigl\langle
         -\frac{p^4}{8m^3}+\frac{\pi\rho}{2m^3}+H_{\rm so}
   \Bigr\rangle_{\big|{R}}\,,
\\[3mm]\displaystyle
\mathcal{E}_{\rm SE}^{(5)}(R) = \alpha^3\frac{4}{3}\left[\ln{\frac{1}{\alpha^2}}-\beta(R)+\frac{5}{6}\right]
       \Bigl\langle Z_1\delta(\mathbf{r}_1)+Z_2\delta(\mathbf{r}_2) \Bigr\rangle_{\big|{R}}\,,
\\[3mm]\displaystyle
\mathcal{E}_{\rm VP}^{(5)}(R) = -\alpha^3\frac{4}{15}
       \Bigl\langle Z_1\delta(\mathbf{r}_1)+Z_2\delta(\mathbf{r}_2) \Bigr\rangle_{\big|{R}}\,.
\end{array}
\end{equation}
Here $\rho=\boldsymbol{\nabla}^2V/(4\pi)$, $H_{\rm so}$ is the electron spin-orbit Hamiltonian (see~\cite{KorobovJPB07} for details), and $\beta(R)$ is the nonrelativistic Bethe logarithm for the bound electron in the two-center problem, whose values as a function of $R$ may be found in the Supplemental Material to Ref.~\cite{Korobov13} or in \cite{Kolos}.

The $m\alpha^7$-order vibrational correction from one-loop self-energy and vacuum polarization is then obtained via the second-order perturbation formalism as
\begin{equation}\label{1loop-vb-a7}
\Delta E_{\rm vb}^{(7)} =
   2\left\langle \chi_{\rm ad} \bigl|
      \mathcal{E}_{\rm BP}^{(4)}(R) Q' (E_0-H_{\rm vb})^{-1} Q' \bigl(\mathcal{E}_{\rm SE}^{(5)}(R)+\mathcal{E}_{\rm VP}^{(5)}(R) \bigr)
   \bigr| \chi_{\rm ad} \right\rangle,
\end{equation}
here $Q'$ is a projection operator onto a subspace orthogonal to $\chi_{\rm ad}(R)$ from Eq.~(\ref{BO_el}). The difference with respect to our previous calculations is that $\mathcal{E}_{\rm SE}^{(5)}(R)$ in Eq.~(\ref{leading}) includes the contribution from the electron anomalous magnetic moment, which was missed in Eq.~(10) of~\cite{Korobov17}.

We now turn to the $m\alpha^8$-order, starting with the one-loop self-energy correction. The higher-order remainder ($m\alpha^8$ and above) for this contribution can be estimated from the hydrogen 1S state results by using the LCAO approximation:
\begin{equation}\label{EffH}
\begin{array}{@{}l}
\displaystyle
\Delta E_{\rm el-1loop}^{(8+)} =
   \alpha^5\left\langle
      \chi_{\rm ad}\bigl|
         \bigl(G_{\rm SE}(1S)-A_{60}\bigr)\Bigl\langle Z_1^3\delta(\mathbf{r}_1)+Z_2^3\delta(\mathbf{r}_2) \Bigr\rangle
         \bigr|\chi_{\rm ad}
   \right\rangle,
\end{array}
\end{equation}
where $G_{\rm SE}(1S)=-30.29024(2)$ is the higher order remainder, and $A_{60}=-30.924149$. Both numbers are taken from Table~II of \cite{YPP19}. This is more accurate than the treatment of Ref.~\cite{Korobov17}, where only the $m\alpha^8$-order term was estimated. The theoretical uncertainty is estimated as equal to $\left| \Delta E_{\rm el-1loop}^{(8+)} - \Delta E_{\rm el-1loop}^{(8log)} \right|$, where $\Delta E_{\rm el-1loop}^{(8log)}$ is the known $m\alpha^8$-order logarithmic term (first term of Eq.~(14) in~\cite{Korobov17}).

The second-order vibrational contribution is expressed
\begin{equation}\label{1loop-vb-a8}
\Delta E_{\rm vb-1loop}^{(8)} =
   2\left\langle \chi_{\rm ad} \bigl|
      \mathcal{E}_{\rm BP}^{(4)}(R) Q' (E_0-H_{\rm vb})^{-1} Q' \bigl(\mathcal{E}_{\rm SE}^{(6)}(R)+\mathcal{E}_{\rm VP}^{(6)}(R) \bigr)
   \bigr| \chi_{\rm ad} \right\rangle,
\end{equation}
where
\[
\begin{array}{@{}l}\displaystyle
\mathcal{E}_{\rm SE}^{(6)}(R) = \alpha^4\left[\frac{139}{32}-2\ln{2}\right]\pi
       \Bigl\langle Z_1^2\delta(\mathbf{r}_1)+Z_2^2\delta(\mathbf{r}_2) \Bigr\rangle_{\big|{R}}\,,
\\[3mm]\displaystyle
\mathcal{E}_{\rm VP}^{(6)}(R) = \alpha^4\frac{5}{48}\pi
       \Bigl\langle Z_1^2\delta(\mathbf{r}_1)+Z_2^2\delta(\mathbf{r}_2) \Bigr\rangle_{\big|{R}}\,.
\end{array}
\]

Finally, we consider two-loop $m\alpha^8$-order corrections. For hydrogen-like atoms, the two-loop correction at orders $m\alpha^8$ and higher is generally expressed in the form~\cite{YPP19,CODATA18}
\begin{equation}
\begin{array}{@{}l}\displaystyle
E_{\rm 2loop}^{(8+)} = \frac{(Z\alpha)^6}{\pi^2 n^3}
   \Bigl[
      B_{63}\ln^3(Z\alpha)^{-2}\!
      +\!B_{62}\ln^2(Z\alpha)^{-2}\!
      +B_{61}\ln(Z\alpha)^{-2}+G^{\rm 2loop}(Z\alpha)
   \Bigr],
\end{array}
\end{equation}
where $G^{\rm 2loop}(Z\alpha)$ is the higher order remainder calculated in \cite{Yerokhin09,Yerokhin18}. We adopt similar notations for hydrogen molecular ions:
\begin{equation} \label{2-loop}
\begin{array}{@{}l}\displaystyle
E_{\rm 2loop}^{(8+)} =
    \frac{\alpha^6}{\pi} \left\langle
      \chi_{\rm ad}\bigl|
               \mathcal{B}_{63}(R) \ln^3(\alpha^{-2}) \!+\! \mathcal{B}_{62}(R) \ln^2(\alpha^{-2})
      \!+\! \mathcal{B}_{61}(R) \ln(\alpha)^{-2} \!+\! G^{\rm 2loop}(1S) \Bigl\langle Z_1^3\delta(\mathbf{r}_1)\!+\!Z_2^3\delta(\mathbf{r}_2) \Bigr\rangle
         \bigr|\chi_{\rm ad}
   \right\rangle.
\end{array}
\end{equation}
Again, the higher-order remainder is estimated using the LCAO approximation. In our calculations, we adopted the value $G^{\rm 2loop}(1S) = -94.5(6.6)$ from~\cite{Karshenboim19}. This is more accurate than our previous treatment~\cite{Korobov17}, where only the $m\alpha^8$-order correction was estimated using $B_{60}(1S)$ instead of $G^{\rm 2loop}(1S)$. The theoretical uncertainty is estimated as equal to the term proportional to $G^{\rm 2loop}(1S)$, after subtraction of the known term of order~$m\alpha^9\ln^2(\alpha)$ (Eq.~(26) of~\cite{YPP19}).

Calculation of the $\mathcal{B}_{6k}(R)$ effective potentials for the two-center problem is described in~\cite{Korobov17}. Since then, a new contribution to the $B_{61}$ coefficient in hydrogen-like atoms from light-by-light scattering diagrams has been found, yielding a correction $B^{\rm LbL}_{61} = 0.830\,309$ for $S$ states~\cite{Szafron19}. For the two-center problem we thus add the following term to $\mathcal{B}_{61}(R)$:
\begin{equation}\label{B61:LbL}
\mathcal{B}_{61}^{\rm LbL}(R) = B^{\rm LbL}_{61} \Bigl\langle Z_1^3\delta(\mathbf{r}_1)+Z_2^3\delta(\mathbf{r}_2) \Bigr\rangle_{\big|{R}}\,.
\end{equation}

The second-order contribution due to vibrational motion is expressed as
\begin{equation}\label{eq:2loop_a8_vb}
\begin{array}{@{}l}
\Delta E_{\rm vb-2loop}^{(8)} =
   2\left\langle \chi_{\rm ad} \bigl|
      \mathcal{E}_{\rm BP}^{(4)}(R) Q' (E_0-H_{\rm vb})^{-1} Q'\mathcal{E}_{\rm 2loop}^{(6)}(R)
   \bigr| \chi_{\rm ad} \right\rangle,
\\[3mm]\displaystyle\hspace{22mm}
   +\left\langle \chi_{\rm ad} \bigl|
      \bigl(\mathcal{E}_{\rm SE}^{(5)}(R)+\mathcal{E}_{\rm VP}^{(5)}(R) \bigr)
      Q' (E_0-H_{\rm vb})^{-1} Q'
      \bigl(\mathcal{E}_{\rm SE}^{(5)}(R)+\mathcal{E}_{\rm VP}^{(5)}(R) \bigr)
   \bigr| \chi_{\rm ad} \right\rangle,
\end{array}
\end{equation}
with
\[
\mathcal{E}_{\rm 2loop}^{(6)}(R) =
   \frac{\alpha^4}{\pi^2}\left[0.538941\right]
         \pi\Bigl\langle Z_1\delta(\mathbf{r}_1)+Z_2\delta(\mathbf{r}_2) \Bigr\rangle_{\big|{R}}\,.
\]
Due to the presence of a logarithmic term in $\mathcal{E}_{\rm SE}^{(5)}(R)$, the vibrational correction is enhanced by a factor of $\ln^2(\alpha^{-2})$ and contributes to the $\mathcal{B}_{62}$, $\mathcal{B}_{61}$ and nonlogarithmic terms. It results in a 1.14~kHz shift for the fundamental vibrational transition, thus the neglect of this term in~\cite{Korobov17} was not justified.

The last corrections requiring new consideration are the muonic and hadronic vacuum polarization corrections. The muonic term is~(see Eq.~(14) of~\cite{YPP19})
\begin{equation}
\mathcal{E}_{\mu{\rm VP}}(R) = \left( \frac{m_e}{m_{\mu}} \right)^2 \mathcal{E}_{\rm VP}^{(5)} (R) \,,
\end{equation}
and the hadronic term may be written as~(Eq.~(15) of~\cite{YPP19})
\begin{equation}
\mathcal{E}_{\rm hadVP}(R) = 0.671(15) \, \mathcal{E}_{\mu{\rm VP}}(R).
\end{equation}
The sum of these two contributions shifts the fundamental transition frequency by 0.25~kHz.

\section{Results} \label{sec4}

Our results for the frequency of the fundamental transition $(L=0,v=0)\to(0,1)$ in HD$^+$ are presented in Table~\ref{ftE} and compared with previous results from Ref.~\cite{Korobov17}. The change in the nonrelativistic transition frequency $\nu_{nr}$ is mainly due to those of the nucleus-to-electron mass ratios (mostly $\mu_p$) and Rydberg constant between the 2014 and 2018 CODATA adjustments. The shift in $\nu_{\alpha^2}$ is due to the nuclear corrections; it stems from the CODATA18 values of the proton and deuteron radii, and from the inclusion of higher-order finite-size and polarizability corrections described in Sec.~\ref{sec2}. The change in $\nu_{\alpha^5}$ comes from the correction of the vibrational contribution [Eq.~(\ref{1loop-vb-a7})]. Finally, several improvements have been made in the calculation of $\nu_{\alpha^6}$ as detailed in Sec.\ref{sec3}: inclusion of vibrational contributions, estimate of the all-order remainder both in one-loop and too-loop corrections, as well as the inclusion of light-by-light scattering diagrams in the two-loop correction. The total shift of the transition frequency is +6.4~kHz, of which +5.4~kHz is due to the shifts in the fundamental constants (+2.7~kHz for $\nu_{nr}$, and +2.7~kHz for the leading-order finite-size correction in $\nu_{\alpha^2}$), and +1.0~kHz comes from the new contributions discussed in Secs.~\ref{sec2} and \ref{sec3}.

The theoretical uncertainty is dominated by the one-loop (Eq.~(\ref{EffH})) and two-loop (term proportional to $G^{\rm 2loop}(1S)$ in Eq.~(\ref{2-loop})) higher-order remainders (see discussion in~\cite{Korobov17}). In the uncertainty from fundamental constants, the largest contribution by far is that of the proton-to-electron mass ratio $\mu_p$ (1.7~kHz and 1.1~kHz using 2014 and 2018 CODATA values, respectively).

\begin{table}[t]
\begin{center}
\caption{Fundamental transition frequency $\nu_{01}$ for the $\mbox{HD}^+$ molecular ion (in kHz). CODATA14 recommended values of fundamental constants were used in~\cite{Korobov17}, and the latest CODATA18 values are used in the present work. Nuclear size and polarizability corrections are included in $\nu_{\alpha^2}$, and ``other'' corrections correspond to the muonic and hadronic vacuum polarization. Theoretical uncertainties of contributions at each order in $\alpha$, if not negligible, are indicated within parentheses. In the final value of the transition frequency, the first error is the theoretical uncertainty, and the second one is due to the uncertainty of fundamental constants.}\label{ftE}
\begin{tabular}{l@{\hspace{16mm}}d@{\hspace{16mm}}d}
\hline\hline
\vrule height 10.5pt width 0pt depth 3.5pt
 & [4] & \hbox{this work} \\
\hline
\vrule height 10pt width 0pt
$\nu_{nr}$       & 57\,349\,439\,952.4    & 57\,349\,439\,955.1   \\
$\nu_{\alpha^2}$ &          958\,151.7    &          958\,154.6   \\
$\nu_{\alpha^3}$ &         -242\,126.3    &         -242\,126.3   \\
$\nu_{\alpha^4}$ &             -1708.9(1) &             -1708.9(1)\\
$\nu_{\alpha^5}$ &               106.4(1) &               105.9(1)\\
$\nu_{\alpha^6}$ &                -2.0(5) &                -0.8(5)\\
other            &                        &                 0.25  \\
\hline
\vrule height 10pt width 0pt
$\nu_{tot}$      & 57\,350\,154\,373.4(0.5)(1.8) & 57\,350\,154\,379.8(0.5)(1.3) \\
\hline\hline
\end{tabular}
\end{center}
\vspace*{-3mm}
\end{table}

Theoretical frequencies for the rovibrational transitions measured in recent experiments are presented in Table~\ref{theor_exp}. They are in very good agreement with experimental results in all cases. For the $(0,0)\!\to\!(1,1)$ transition the combined uncertainty $u=0.86$ kHz ($u^{\rm exp}=0.16$ kHz and $u^{\rm theor,spin}=0.85$ kHz) is given, for the $v=(3,0)\!\to\!(3,9)$ transition, the revised experimental value from~\cite{Germann21} is used. Numerical results of calculations for all the contributions considered in Sec.~\ref{sec3} (with the $m\alpha^6$-order relativistic correction as well) for a wide range of rovibrational states are given in the Supplemental Material~\cite{SM}.

\begin{table}[h]
\begin{center}
\caption{Theoretical and experimental spin-averaged transition frequencies (in kHz). CODATA18 values of fundamental constants were used in the calculations. For theoretical values, the first uncertainty is due to yet uncalculated terms and used approximations in theory, while the second uncertainty is due to inaccuracy in the CODATA18 recommended mass values.}\label{theor_exp}
\begin{tabular}{c@{\hspace{8mm}}c@{\hspace{8mm}}c}
\hline\hline
$(L,v)\to(L',v')$ & theory & experiment \\
\hline
$(0,0)\to(1,0)$ &  1\,314\,925\,752.932(19)(61)  & 1\,314\,925\,752.910(17) \\
$(0,0)\to(1,1)$ & 58\,605\,052\,163.9(0.5)(1.3)  & 58\,605\,052\,164.24(86) \\
$(3,0)\to(3,9)$ & 415\,264\,925\,502.8(3.3)(6.7) & 415\,264\,925\,501.8(1.3) \\
\hline\hline
\end{tabular}
\end{center}
\end{table}

Using these improved theoretical predictions, we give in Table~\ref{mupe} some updated determinations of the proton-to-electron mass ratio from HD$^+$ spectroscopy. We follow the least-squares fitting procedure used in the CODATA adjustments and described in Appendix~E of~\cite{CODATA98}. The dependence of HD$^+$ transition frequencies on fundamental constants is linearized using a first-order Taylor expansion around their starting (CODATA18) values; first-derivative coefficients are obtained by the methods described in~\cite{Schiller05,Karr16}. Following the CODATA approach, we adjust the individual particle masses rather than mass ratios. In more detail, we use CODATA18 values of $m_d$, $R_{\infty}$, $r_d$, $r_p$, include as an additional data point the latest measurement of the electron mass~\cite{Kohler15}, and solve for the electron and proton masses. $m_p/m_e$ and its uncertainty are then deduced from the adjusted values of $m_p$ and $m_e$, taking the correlation between them into account. The covariance matrix of the HD$^+$ input data (see~\cite{CODATA98} for definition) is built including the three (uncorrelated) sources of uncertainties: experimental, theoretical, and parametric, the latter stemming from uncertainties of fundamental constants that are not directly involved in the adjustment ($m_d$, $R_{\infty}$, $r_d$, $r_p$). When several HD$^+$ measurements are combined, the following assumptions are made regarding correlations: (i) uncorrelated experimental uncertainties, (ii) fully correlated theoretical uncertainties, and (iii) correlations between parametric uncertainties are included taking into account the correlation coefficients between fundamental constants available from~\cite{CODATA-web}.

Values obtained in this way from each single HD$^+$ experiment are given in the first three lines of Table~\ref{mupe}, and the combined result from the three measurements in the fourth line. Note that its uncertainty is only slightly reduced with respect to that of individual lines due to strong correlation between them. The latter value, although slightly higher (by 1.4 combined standard deviations), is in good agreement with that obtained from recent high-precision mass spectroscopy measurements of $m_p$~\cite{Heisse19}, $m_d$~\cite{Rau20}, $m($HD$^+)$~\cite{Rau20}, and $m_d/m_p$~\cite{Fink20} in Penning traps: $m_p/m_e = 1836.152\,673\,343(60)$~\cite{Kohler15,Rau20}.

Finally, the HD$^+$ data can be combined with the mass spectrometry measurements and the CODATA18 values of $R_{\infty}$, $r_d$, $r_p$ to simultaneously determine the three particle masses, $m_e$, $m_p$ and $m_d$. The mass ratios $m_p/m_e$ and $m_d/m_p$ deduced from this adjustment are shown in the last line of Table~\ref{mupe}. Relative uncertainties of $1.8 \times 10^{-11}$ (for $m_p/m_e$) and $1.6 \times 10^{-11}$ (for $m_d/m_p$) are obtained, improved by factors of 3.3 and 3.5 with respect to CODATA18.

\begin{table}[h]
\begin{center}
\caption{Determinations of mass ratios using HD$^+$ spectroscopy.}\label{mupe}
\begin{tabular}{c@{\hspace{8mm}}c@{\hspace{8mm}}c}
\hline\hline
data & $m_p/m_e$ & $m_d/m_p$ \\
\hline
$(0,0)\to(1,0)$ & 1836.152\,673\,480(63) & \\
$(0,0)\to(1,1)$ & 1836.152\,673\,40(10)~ & \\
$(3,0)\to(3,9)$ & 1836.152\,673\,457(73) & \\
\hline
HD$^+$          & 1846.152\,673\,466(61) & \\
\hline
HD$^+$/Penning  & 1836.152\,673\,454(33) & 1.999\,007\,501\,243(31) \\
\hline\hline
\end{tabular}
\end{center}
\end{table}

In conclusion, we have presented a revised and improved theory of spin-averaged transition frequencies in hydrogen molecular ions. Further progress in precision now requires calculations of nonlogarithmic $m\alpha^8$-order one- and two-loop corrections.

\emph{Acknowledgements.} The authors thank J.C.J.~Koelemeij for his help with the least-squares adjustments of fundamental constants. V.I.K. acknowledges support of the Russian Foundation for Basic Research under Grant No.~19-02-00058-a.


\begin{thebibliography}{99}

\bibitem{Schiller20} S.~Alighanbari, G.S.~Giri, F.L.~Constantin, V.I.~Korobov, and S.~Schiller,
  Precise test of quantum electrodynamics and determination of fundamental constants with HD$^+$ ions.
  Nature \textbf{581}, 152 (2020).

\bibitem{Patra20} S.~Patra, M.~Germann, J.-Ph.~Karr, M.~Haidar, L.~Hilico, V.I.~Korobov, F.M.J.~Cozijn, K.S.E.~Eikema, W.~Ubachs, and J.C.J.~Koelemeij,
  Proton-electron mass ratio from laser spectroscopy of HD$^+$ at the part-per-trillion level.
  Science \textbf{369}, 1238 (2020).

\bibitem{Kortunov21} I.~Kortunov, S.~Alighanbari, M.G.~Hansen, G.S.~Giri, S.~Schiller, and V.I.~Korobov,
  Proton-electron mass ratio by high-resolution optical spectroscopy of ion ensemble in the resolved-carrier regime.
  Nature Phys.\ \textbf{17}, 569 (2021).

\bibitem{Korobov17} V.I.~Korobov, L.~Hilico, and J.-Ph.~Karr,
  Fundamental transitions and ionization energies of the hydrogen molecular ions with few ppt uncertainty.
  Phys.\ Rev.\ Lett.\ \textbf{118}, 233001 (2017).

\bibitem{Germann21} M.~Germann, S.~Patra, J.-Ph.~Karr, L.~Hilico, V.I.~Korobov, E.J.~Salumbides, K.S.E.~Eikema, W.~Ubachs, and J.C.J.~Koelemeij,
  Three-body QED test and fifth-force constraint from vibrations and rotations of HD$^+$.
  Phys.\ Rev.\ Research \textbf{3}, L022028 (2021).

\bibitem{Zhong15} Z.-X.~Zhong, X.~Tong, Z.-C.~Yan, and T.-Y.~Shi,
  High-precision spectroscopy of hydrogen molecular ions.
  Chin.\ Phys.~B \textbf{24}, 053102 (2015).

\bibitem{Schmidt20} J.~Schmidt, T.~Louvradoux, J.~Heinrich, N.~Sillitoe, M.~Simpson, J.-Ph.~Karr, and L.~Hilico,
  Trapping, Cooling, and Photodissociation Analysis of State-Selected H$_2^+$ Ions Produced by (3 + 1) Multiphoton Ionization.
  Phys.\ Rev.\ Appl.\ \textbf{14}, 024053 (2020).

\bibitem{Schiller14} S.~Schiller, D.~Bakalov, and V.I.~Korobov,
  Simplest Molecules as Candidates for Precise Optical Clocks.
  Phys.\ Rev.\ Lett.\ \textbf{113}, 023004 (2014).

\bibitem{Karr14} J.-Ph.~Karr,
  H$_2^+$ and HD$^+$: Candidates for a molecular clock.
  J.\ Mol.\ Spectrosc.\ \textbf{300}, 37 (2014).

\bibitem{Schmidt05} P.O.~Schmidt, T.~Rosenband, C.~Langer, W.M.~Itano, J.C.~Bergquist, and D.J.~Wineland,
  Spectroscopy using quantum logic.
  Science \textbf{309}, 749 (2005).

\bibitem{Wolf16} F. Wolf, Y.~Wan, J.C.~Heip, F.~Gebert, C.~Shi, and P.O.~Schmidt,
  Non-destructive state detection for quantum logic spectroscopy of molecular ions.
  Nature \textbf{530}, 457 (2016).

\bibitem{Chou20} C.W.~Chou, A.L.~Collopy, C.~Kurz, Y.~Lin, M.E.~Harding, P.N.~Plessow, T.~Fortier, S.~Diddams, D.~Leibfried, and D.R.~Leibrandt,
  Frequency-comb spectroscopy on pure quantum states of a single molecular ion.
  Science \textbf{367}, 1458 (2020).

\bibitem{YPP19} V.A.~Yerokhin, K.~Pachucki, and V.~Patk\'o\v{s},
  Theory of the Lamb Shift in Hydrogen and Light Hydrogen-Like Ions.
  Ann.\ Phys.\ (Berlin) \textbf{531}, 1800324 (2019).

\bibitem{Szafron19} R.~Szafron, E.Yu.~Korzinin, V.A.~Shelyuto, V.G.~Ivanov, and S.G.~Karshenboim,
  Virtual Delbr\"uck scattering and the Lamb shift in light hydrogenlike atoms.
  Phys.\ Rev.~A \textbf{100}, 032507 (2019).

\bibitem{Karshenboim19} S.G.~Karshenboim, A.~Ozawa, V.A.~Shelyuto, R.~Szafron, and V.G.~Ivanov,
  The Lamb shift of the 1sstate in hydrogen: Two-loop and three-loop contributions.
  Phys.\ Lett.~B \textbf{795}, 432 (2019)

\bibitem{CODATA14} P.J.~Mohr, B.N.~Taylor, and D.B.~Newell,
  CODATA recommended values of the fundamental physical constants: 2014.
  Rev.\ Mod.\ Phys.\ \textbf{88}, 035009 (2016).

\bibitem{CODATA18} E.~Tiesinga, P.J.~Mohr, D.B.~Newell, and B.N.~Taylor,
  CODATA recommended values of the fundamental physical constants: 2018.
  Rev.\ Mod.\ Phys.\ \textbf{93}, 025010 (2021).

\bibitem{Yerokhin09} V.A.~Yerokhin,
  Two-loop self-energy for the ground state of medium-$Z$ hydrogenlike ions.
  Phys.\ Rev.~A \textbf{80}, 040501(R) (2009).

\bibitem{Yerokhin18} V.A.~Yerokhin,
  Two-loop self-energy in the Lamb shift of the ground and excited states of hydrogenlike ions.
  Phys.\ Rev.~A \textbf{97}, 052509 (2018).

\bibitem{Korobov06} V.I.~Korobov,
  Leading-order relativistic and radiative corrections to the rovibrational spectrum of H$_2^+$ and HD$^+$ molecular ions,
  Phys.\ Rev.~A \textbf{74}, 052506 (2006).

\bibitem{Friar97} J.L.~Friar, G.L.~Payne,
  Higher-order nuclear-polarizability corrections in atomic hydrogen.
  Phys.\ Rev.~C \textbf{56}, 619 (1997).

\bibitem{Friar97b} J.L.~Friar, G.L.~Payne,
  Higher-order nuclear-size corrections in atomic hydrogen.
  Phys.\ Rev.~A \textbf{56}, 5173 (1997).

\bibitem{Korobov12} V.I.~Korobov and Z.-X.~Zhong,
  Bethe logarithm for the H$_2^+$ and HD$^+$ molecular ions,
  Phys.\ Rev.~A \textbf{86}, 044501 (2012).

\bibitem{KorobovJPB07} V.I.~Korobov and Ts.~Tsogbayar,
  Relativistic corrections of order $m\alpha^6$ to the two-centre problem.
  J.~Phys.~B:\ At.\ Mol.\ Opt.\ Phys.\ \textbf{40}, 2661 (2007).

\bibitem{Wolniewicz80} L.~Wolniewicz and J.D.~Poll,
  The vibration-rotational energies of the hydrogen molecular ion HD$^+$.
   J.~Chem.\ Phys.\ \textbf{73}, 6225 (1980).

\bibitem{Carrington89} A.~Carrington, I.R. McNab, and Ch.A.~Montgomerie,
  Spectroscopy of the hydrogen molecular ion.
  J.~Phys.~B:\ At.\ Mol.\ Opt.\ Phys.\ \textbf{22}, 3551 (1989).

\bibitem{JCP_PRL05} A.~Czarnecki, U.D.~Jentschura, and K.~Pachucki,
  Calculation of the One- and Two-Loop Lamb Shift for Arbitrary Excited Hydrogenic States.
  Phys.\ Rev.\ Lett.\ \textbf{95}, 180404 (2005);

\bibitem{JCP_PRA05} U.D.~Jentschura, A.~Czarnecki, and K.~Pachucki,
  Nonrelativistic QED approach to the Lamb shift.
  Phys.\ Rev.~A \textbf{72}, 062102 (2005).

\bibitem{Korobov13} V.I.~Korobov, L.~Hilico, and J.-Ph.~Karr,
  Calculation of the relativistic Bethe logarithm in the two-center problem.
  Phys.\ Rev.~A \textbf{87}, 062506 (2013).

\bibitem{KorobovPRL14} V.I.~Korobov, L.~Hilico, and J.-Ph.~Karr,
  $m\alpha^7$-Order Corrections in the Hydrogen Molecular Ions and Antiprotonic Helium.
  Phys.\ Rev.\ Lett.\ \textbf{112}, 103003 (2014).

\bibitem{KorobovPRA14} V.I.~Korobov, L.~Hilico, and J.-Ph.~Karr,
  Theoretical transition frequencies beyond 0.1 ppb accuracy in H$_2^+$, HD$^+$, and antiprotonic helium.
  Phys.\ Rev.~A \textbf{89}, 032511 (2014).

\bibitem{KarrVP17} J.-Ph.~Karr, L.~Hilico, and V.I.~Korobov,
  One-loop vacuum polarization at $m\alpha^7$ and higher orders for three-body molecular system.
  Phys.\ Rev.~A \textbf{95}, 042514 (2017).

\bibitem{Kolos} R.~Bukowski, B.~Jeziorski, R.~Moszy\'nski, and W.~Ko\l os,
  Bethe logarithm and Lamb Shift for the Hydrogen Molecular Ion.
  Int.\ J.~Quantum Chem.\ \textbf{42}, 287 (1992).

\bibitem{SM} See Supplemental Material for a complete set of Tables with various contributions to the energies of a wide range of rovibrational states in HD$^+$ ion.

\bibitem{CODATA98} P.J.~Mohr and B.N.~Taylor,
  CODATA recommended values of the fundamental physical constants: 1998.
  Rev.\ Mod.\ Phys.\ \textbf{72}, 351 (2000).

\bibitem{Schiller05} S.~Schiller and V.I.~Korobov,
  Tests of time independence of the electron and nuclear masses with ultracold molecules.
  Phys.\ Rev.~A \textbf{71}, 032505 (2005).

\bibitem{Karr16} J.-Ph.~Karr, L.~Hilico, J.C.J.~Koelemeij, and V.I.~Korobov,
  Hydrogen molecular ions for improved determination of fundamental constants.
  Phys.\ Rev.~A \textbf{94}, 050501(R) (2016).

\bibitem{Kohler15} F.~K\"ohler, S.~Sturm, A.~Kracke, G.~Werth, W.~Quint, and K.~Blaum,
  The electron mass from $g$-factor measurements on hydrogen-like carbon $^{12}$C$^{5+}$.
  J.\ Phys.~B: At.\ Mol.\ Opt.\ Phys.\ \textbf{48}, 144032 (2015).

\bibitem{CODATA-web} http://physics.nist.gov/cuu/Constants/

\bibitem{Heisse19} F.~Hei{\ss}e, S.~Rau, F.~K\"ohler-Langes, W.~Quint, G.~Werth, S.~Sturm, and K.~Blaum,
  High-precision mass spectrometer for light ions.
  Phys.\ Rev.\ A \textbf{100}, 022518 (2019).

\bibitem{Rau20} S.~Rau, F.~Hei{\ss}e, F.~K\"ohler-Langes, S.~Sasidharan, R.~Haas, D.~Renisch, C.E.~D\"ullmann, W.~Quint, S.~Sturm, and K.~Blaum,
  Penning trap mass measurements of the deuteron and the HD$^+$ molecular ion.
  Nature \textbf{585}, 43 (2020).

\bibitem{Fink20} D.J.~Fink and E.G.~~Myers,
  Deuteron-to-Proton Mass Ratio from the Cyclotron Frequency Ratio of H$_2^+$ to D$^+$ with H$_2^+$ in a Resolved Vibrational State.
  Phys.\ Rev.\ Lett.\ \textbf{124}, 013001 (2020).

\end{thebibliography}
\end{document}